\documentclass[a4paper,12pt]{article}
\usepackage[utf8]{inputenc}

\title{Professor "Nonlinear"\\
Obituary and memoir for Roman Juszkiewicz, 1952-2012\footnote{Roman Juszkiewicz passed away on 28th of January 2012}}
\author{Wojciech A. Hellwing\footnote{pchela@icm.edu.pl}}

\begin{document}

\maketitle

\begin{abstract}
On 28th and 29th of January 2013 we held an international meeting in Zielona G\'ora (Poland)
honouring the first anniversary of premature passing away of Professor Roman Juszkiewicz.
We have celebrated an opening of a new seminar room at the University of
Zielona G\'ora commemorated to the memory of Roman Juszkiewicz and we have shared
our anecdotes and memories of this great scientist and friend. Here we want to present
a limited and short memoir and obituary for Roman Juszkiewicz.
\end{abstract}

Professor "Nonlinear" is the honorary nickname which Bernard Jones gave 
to Roman in his summary of the international cosmology workshop in July 2011 
at the Copernicus Center in Warsaw. Roman Juszkiewicz had conquered a special 
position in our hearts, of which the meeting in Warsaw was a special 
and telling testimony. Over the year 2010 Roman had run into a serious health 
problem that prevented him from travelling. As the meeting was specially 
organized for him, as a welcome back gesture, we decided that if Roman 
cannot make it to the meeting elsewhere we take the meeting to him. It was 
not to be, we all know that by July 2011, the sudden deterioration of his 
health prevented him from participation. 

The Professor Nonlinear distinction did not come by coincidence. When we 
look at the impressive scientific record of Roman Juszkiewicz, we may
immediately 
appreciate that his honorary name is well deserved. Its  history starts 
with the very first paper that Roman submitted to Monthly Notices of 
the Royal Astronomical Society, in 1981, and that marked the start of his international 
scientific career. The paper contained a thorough mathematical and physical 
perturbation analysis in the weakly nonlinear limit for the equations 
describing the evolution of large scale structures in the Universe. 
Even today, more than 30 years later, the paper is still cited. 

Those who knew Roman at that time recall that he had decided to 
extract as much information as possible from the fact that the 
universe is nonlinear, an observation to which he kept on drawing 
special attention throughout all his career. More than anything he, 
together with an impressive array of students he guided along 
these lines, succeeded in this this self-imposed task. He contributed 
many solutions and results to modern cosmology from the perturbation 
analysis of weakly nonlinear cosmic fields. Of course these were 
not his only achievements, he touched many areas in cosmology. 

To appreciate better the many accomplishments that Roman attained, 
we should try to know and understand him as a person. Roman 
never undertook anything half-heartedly. Taking half measures 
was not acceptable to him, and this striving for perfection was 
at the core of every aspect of his life. Before anything else 
his research, as we know that the vast majority of his scientific 
publications radiate excellence and have become contributions 
with a high number of citations. But his friends and peers also 
know Roman as a true "bon vivant", who applied the same standards 
of perfection to his love of food, wine and every aspect of culture. 

Roman was a student of the great and famous Jakov Zel'dovich. Zel'dovich 
supervised his master thesis at Lomosonov University in Moskow. In 1981
he obtain his PhD degree in physics at the University of Warsaw.  
Roman's career quickly gained pace, attested by the fact that the ten 
years following his PhD he spent working in prestigious international 
institutes. His many collaborators included, amongst others, 
John Barrow, Joe Silk, Jim Peebles, Marc Davis, Jerry Ostriker, 
Francois Bouchet, Edmund Bertschinger and Rien van de Weygaert. 
It would take too much 
space to list the complete list of his collaborators on this page. 
In the UK he worked at the IOA in Cambridge and in Sussex, 
and later he worked in the United States at Berkeley and in Princeton at 
the Institute for Advanced Studies. Before he returned to Poland, 
in the second half of the nineties, he was a senior researcher in 
well-known European institutes, such as the University of Geneva 
and the Institut d'Astrophysique in Paris. In 1996 he assumed a 
professor position at the Nicolaus Copernicus Astronomical Center 
in Warsaw. Five years later, in 2001, he also became professor 
at the newly founded University of Zielona Gora. In these years 
he became the mentor and supervisor of many young researchers 
from Poland and France. 

During his successful - be it too short - career he became known 
for a range of important and ground-breaking results. Some of these 
I want to draw attention to. Very important was his work with 
Ed Bertschinger and Jim Peebles at the end of the eighties that 
demonstrated beyond doubt that the standard cosmological model 
in those days - the so-called SCDM or "standard cold dark matter 
model" - would lead to some serious discrepancies with astronomical 
observations. Several years later the SCDM model indeed got set 
aside and in the meantime has been replaced by the LCDM 
"Lambda Cold Dark Matter" model. Perhaps his greatest acclaim 
Roman achieved with his work in formulating and developing the 
theory of nonlinear gravitational instability. One may even say 
that he set up a school of highly reputable young French 
cosmologists known for their major contributions in this field, 
with names such as Francois Bouchet, Francis Bernardeau and 
Stephane Colombi. The results of this mathematical machinery 
have been crucial in underpinning our understanding of the 
onset of nonlinearity in the evolving large scale structure, 
something which hitherto was only possible within the framework 
of N-body simulations. Roman also applied perturbation theory to the 
study of large-scale velocity flows of galaxies, offering a 
direct means of "weighing" the Universe. Together with Hume 
Feldman and other collaborators, in 2003 he published a study 
in which they had managed to estimate the total density of 
non-relativistic matter in the Universe. The value they found 
was consistent with estimates that other studies had made. However, 
the importance of the determination by Feldman, Juszkiewicz et al. 
was that it offered an entirely independent means for 
estimating this key cosmological parameter. It included and 
involved physical phenomena that had not before been used and 
that allowed them to work out the estimates independent of 
intermediate assumptions or priors. In addition to these contributions 
professor Juszkiewicz was also involved in a large range of other 
studies and scientific interests. He worked on the cosmic microwave
background radiation, on the statistics of high order correlations 
of cosmic fields, on the topology of the Universe and also
on modified gravity models. In all of these subjects, he made 
major contributions. 

Not only was Roman a great scientist. To all of us he is known 
for his cordial and warm social personality. Like a butterfly, 
whenever he travelled he made new friends and established 
strong friendships. He had a gift that allowed him to quickly win 
over people and make long lasting friendships. I think I do not 
exaggerate when telling that all of his friends do not only 
know him as expert in elegant mathematics, but also as a world 
expert of exquisite food and the best of wines. Amongst his 
friends, it was often joked that Roman would manage in every 
new city he visited to discover new charming cafes and fabulous 
restaurants of which the local astronomer and cosmologist would 
not yet have been aware. Roman loved to hang out and work in those 
places. 

By losing Roman we not only miss a great scientist, but also a great 
friend. A warm and friendly man, who constantly amazed us with his new 
ideas, and entertain us with his many and diverse stories. Aside from this,
we lose a man who worshipped life in every aspect, and who always 
prepared to share his positive energy, no matter the
circumstances. It is this image of Professor Nonlinear that we 
will keep in our hearts and memories ... 

\section*{Acknowledgements}
WAH is very grateful to Rien van de Weyagert for carefully reading the
manuscript and providing many useful suggestion, which in the end
help improve the quality of the text significantly. Furthermore
Staszek Bajtlik, Bernard T. Jones, Sergei Shandarin, Joe Silk, Rien van de Weygaert,
Miros\l aw Panek, Adi Nusser, Ruth Durrer, Andrew H. Jaffe, Carlos S. Frenk and Hume Feldman
are acknowledged for sharing stories and anecdotes from Roman's life.
\end{document}